\documentstyle[twocolumn,prb,aps,epsfig]{revtex}

\begin{document}
\title{Modeling the electronic behavior of $\gamma$-LiV$_2$O$_5$: a 
 microscopic study.}

\author{Roser Valent\'\i$^1$, T. Saha-Dasgupta$^2$,
      J.V. Alvarez$^1$, K. Po\v{z}gaj\v{c}i\'{c}$^1$ and Claudius Gros$^1$ 
       } 

\address{$^1$Fakult\"at 7, Theoretische Physik,
 University of the Saarland,
66041 Saarbr\"ucken, Germany.}

\address{$^2$ Bose National Centre for Basic Sciences, JD Block, Sector 3,
 Salt Lake City, Kolkata 700098, India.}

\date{\today}
\maketitle

\begin{abstract}
We determine the electronic structure of 
the one-dimensional spin-$\frac{1}{2}$ Heisenberg compound
$\gamma$-LiV$_2$O$_5$, which
has two inequivalent vanadium ions, V(1) and V(2), via
density-functional calculations. We find a relative V(1)-V(2) charge
ordering of roughly $70:30$.
We discuss the influence of the charge ordering on
the electronic structure and the magnetic behavior. We
give estimates of the basic hopping matrix elements and
compare with the most studied
$\alpha '$-NaV$_2$O$_5$.

\end{abstract}

%%%%%%%%%%%%%%%%%%%%%%%%%%%%%%%%%%%%%%%%%%%%%%%%%%%%%%%%%%%%%%%%%%%
%%%%%%%%%%%%%%%%%%%%%%%%%%%%%%%%%%%%%%%%%%%%%%%%%%%%%%%%%%%%%%%%%%%

\vspace*{1cm}

Low-dimensional transition metal compounds have been intensively
studied in the past years. Among those,  the
quarter-filled ladder compound \cite{Smolinski98,Horsch98} 
$\alpha '$-NaV$_2$O$_5$ has become a model substance
for the study of spin-charge and orbital coupling.  
The  coupling between spin and orbital ordering is a central issue in the
somewhat more complex colossal magnetoresistance  
(CMR)-materials \cite{CMR}, the advantage of
 $\alpha '$-NaV$_2$O$_5$ as a model compound is the spatial
separation of the two $d_{xy}$-orbitals onto two 
spatially separated V$^{4.5+}$-sites, arranged in ladders.
These two orbitals undergo a charge order transition \cite{Ohama99} at
$T_c=34\,\mbox{K}$ with  
2V$^{4.5+}$$\rightarrow$V$^{4+}$+V$^{5+}$.
Simultaneously to the charge order and the respective
lattice ditortions \cite{crystal},  a spin-gap \cite{Isobe96} opens,
indicating substantial spin-charge coupling in this compound. The
physics of this transition is being studied intensively
\cite{Gros_et_al,Bernet_et_al}. 

A much less studied, though not less intriguing system belonging
to the same vanadium oxide family is $\gamma$-LiV$_2$O$_5$.
Susceptibility measurements \cite{Isobe_96_2},
as well as NMR experiments \cite{Fujiwara97},
on this compound suggest a one-dimensional spin-$\frac{1}{2}$
 Heisenberg-like
behavior and
there is no indication of a phase transition at 
lower temperatures. To our knowledge, 
there is no microscopic study of the electronic
structure of this material discussing the magnetic 
interactions responsible for such behavior. 
We present here a density-functional analysis (DFT) 
of this system and
calculate the possible exchange matrix elements via the 
Linear Muffin-Tin Orbital
(LMTO) based downfolding method \cite{Andersen00}
and a tight-binding model. 

$\gamma$-LiV$_2$O$_5$ offers the possibility, 
due to its close relation to $\alpha '$- NaV$_2$O$_5$,
to study the influence of charge ordering on 
the electronic structure. In particular this opens
the question on how far the charge ordering and
the corresponing crystallographic distortions
alter the magnetic interactions.  We discuss several possible
scenarios compatible with the experimental susceptibility for
$\gamma$-LiV$_2$O$_5$ for the underlying magnetic model
(i) A zig-zag chain model of V(1)-ions,
(ii) A  double-chain model of V(1)-ions, 
(iii) An asymmetric quarter-filled ladder model.

%%%%%%%%%%%%%%%%%%%%%%%%%%%%%%%%%%%%%%%%%%%%%%%%%%%%%%%%%%%%%%%%%%%
%%%%%%%%%%%%%%%%%%%%%%%%%%%%%%%%%%%%%%%%%%%%%%%%%%%%%%%%%%%%%%%%%%%

{\it {Crystal structure.-}}
$\gamma$-LiV$_2$O$_5$ has a layered structure of VO$_5$
square pyramids with lithium ions between
the layers. It crystallizes \cite{Galy_55}
in the orthorhombic centrosymmetric
space group $D^{16}_{2h}-Pnma$ and has two crystallographic
inequivalent vanadium sites, V(1) and V(2),
which form two different zig-zag chains running along the $y$
axis. Within the layers, V(1)O$_5$ zig-zag chains are linked
to V(2)O$_5$ zig-zag chains by corner sharing via
the bridging O(1). The existence of two types of V-sites has been also 
verified by NMR experiments \cite{Fujiwara_98}.

\begin{figure}[t]
(a)

\vspace*{5pt}

\centerline{
\epsfig{file=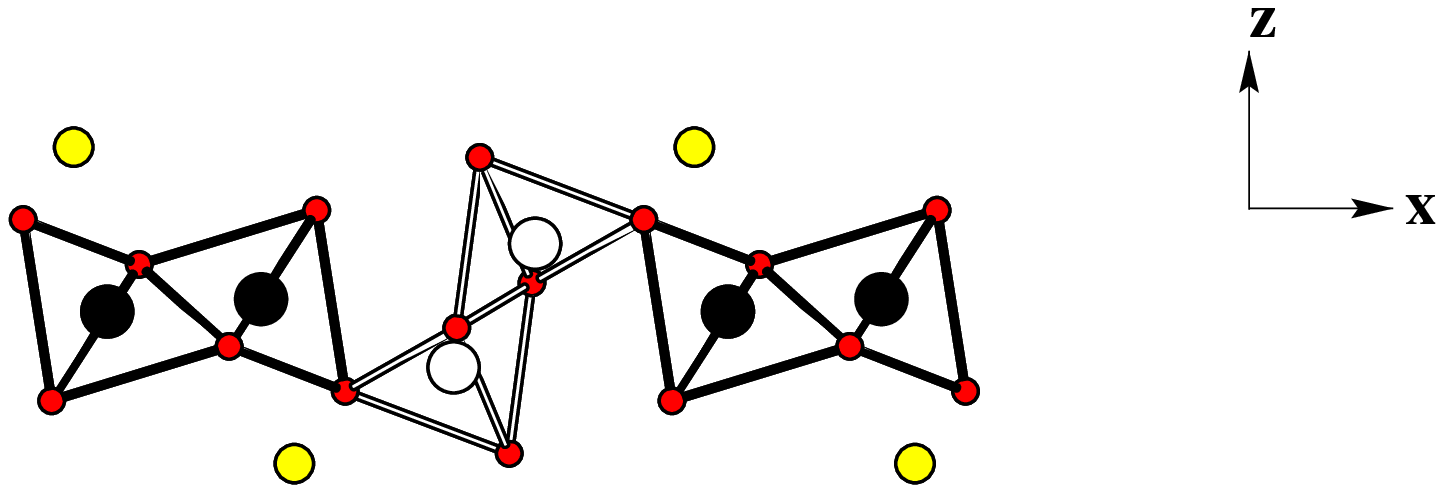,width=0.44\textwidth}
           } 

\vspace{-20pt}
(b) \vspace{-30pt}

\centerline{
\epsfig{file=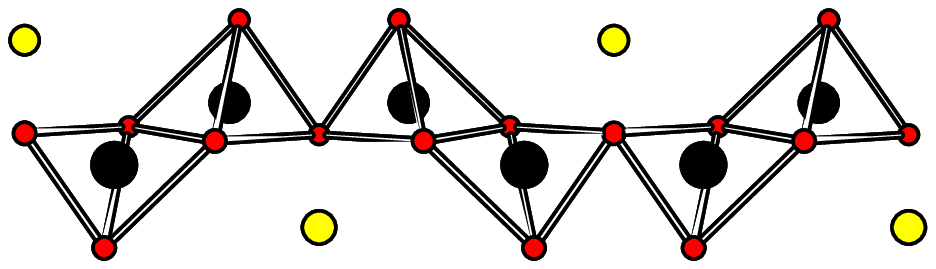,width=0.18\textwidth,angle=-90}
           }
\vspace{0pt}
\caption{\label{struc_xz}
Crystal structure of (a) $\gamma$-LiV$_2$O$_5$ and 
(b) $\alpha '$-NaV$_2$O$_5$ projected in $(xz)$ plane. 
 For $\gamma$-LiV$_2$O$_5$ the $(xz)$ cut through one
of the
 two equivalent $xy$ planes is shown here. The large
circles are the V-ions, black and white for V(1) and V(2) respectively
in $\gamma$-LiV$_2$O$_5$ and black for $\alpha '$-NaV$_2$O$_5$.
The oxygens are represented by the smaller
circles. The alkali-ions (Li,Na),
shown by grey circles, are located in between the planes, close to the
bridging oxygens.}
\end{figure}

In Fig.\ \ref{struc_xz} we show the crystal structure of
 $\gamma$-LiV$_2$O$_5$ and $\alpha'$-NaV$_2$O$_5$ projected
on the $xz$ plane.   The
angle between the basal plane of the (nearly)
square V(1)O$_5$/ V(2)O$_5$ pyramids and the $x$-axis
is about $+30^\circ$/ $-30^\circ$ respectively for $\gamma$-LiV$_2$O$_5$.
The basal plane of the VO$_5$ pyramids in
$\alpha'$-NaV$_2$O$_5$ is, on the other hand, nearly
parallel to the $x$-axis.
 Note that the VO$_5$ square pyramids
are oriented along $x$ as down-down-up-up 
in $\alpha'$-NaV$_2$O$_5$ while in 
$\gamma$-LiV$_2$O$_5$ the orientation 
is down-up-down-up. 

\begin{figure}[t]
\centerline{
\epsfig{file=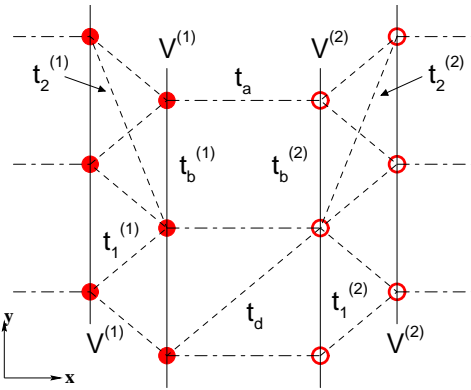,width=0.35\textwidth} %,angle=-90}
           }
\vspace{4pt}
\caption{\label{band_Li}
Hopping parameters used for the tight-binding model
for $\gamma$-LiV$_2$O$_5$. The arrangement of the
V-ions in an $xy$-plane is topologically identical to
the one in $\alpha '$-NaV$_2$O$_5$, i.e.\ they form a Trellis lattice.
The $t_d$ and the $t_2^{(1/2)}$ 
are shown only partially. Not shown are the onsite energies 
$\pm\varepsilon_0$ of the $V(1/2)$ sites.}
\end{figure}

          From the structural analysis
it has been proposed \cite{Galy_55} that the oxidations of V(1) and V(2) are, 
respectively, V$^{4+}$ and V$^{5+}$.
The temperature dependence of the susceptibility $\chi (T)$  
follows that of a  one-dimensional spin-$\frac{1}{2}$ Heisenberg model 
with an exchange interaction of $J_{exp}=308\,\mbox{K}$ and a 
gyromagnetic factor $g=1.8$ \cite{Isobe_96_2}.   
Note that depending on the magnitude of the exchange couplings, 
J$_1$ -between V(1)-ions in edge-shared pyramids- and 
J$_b$ -between V(1)-ions in corner-shared pyramids- 
along $y$ the system can be treated
as Heisenberg zig-zag chains (J$_1$ $\gg$ J$_b$) or
as Heisenberg double-linear chains (J$_1$ $\ll$ J$_b$),
compare with Fig.\ \ref{band_Li}.

A third possible interpretation of the nature of
$\gamma$-LiV$_2$O$_5$ compatible with the 
experimental susceptibility relies on the possibility
of a {\it partially} charge-ordered
system i.e.\ V(1)-sites somewhat closer to V$^{4+}$-oxidation
and V(2)-sites closer to  V$^{5+}$-oxidation.  A picture
of an asymmetric ladder with one electron per V(1)-O-V(2) rung 
would then describe the system in analogy to
$\alpha'$-NaV$_2$O$_5$  where the magnetic
interactions  among the constituent ladders are
weak \cite{Gros99}. In the following, 
we will investigate these three scenarios.

%%%%%%%%%%%%%%%%%%%%%%%%%%%%%%%%%%%%%%%%%%%%%%%%%%%%%%%%%%%%%5
%%%%%%%%%%%%%%%%%%%%%%%%%%%%%%%%%%%%%%%%%%%%%%%%%%%%%%%%%%%%%5

\begin{figure}[t]
%\centerline{
%\epsfig{file=LiV2O5_spag1.ps,width=0.50\textwidth,
%                            height=0.4\textwidth}}
%LiV2O5_spag1.ps
\epsfxsize=0.5\textwidth
\centerline{\epsffile{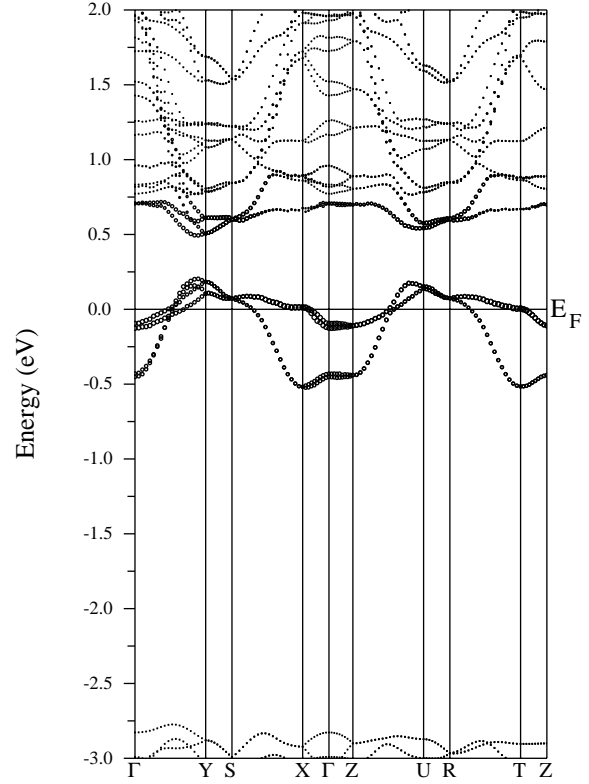}\hfill}

\vspace{4pt}
\caption{\label{Li_spag}
LDA results for LiV$_2$O$_5$. The path is 
along $\Gamma$=(0,0,0), Y=(0,$\pi$,0),
S=($\pi$,$\pi$,0),
X=($\pi$,0,0), $\Gamma$,
Z=(0,0,$\pi$),
U=(0,$\pi$,$\pi$),
R=($\pi$,$\pi$,$\pi$),
T=($\pi$,0,$\pi$), Z.
The V(1)-$3d_{xy}$ character of the bands is shown with bigger circles.}
\end{figure}

 {\it  Band structure.-} 
We have calculated the energy bands (see Fig.\ \ref{Li_spag})
of $\gamma$-LiV$_2$O$_5$ within
DFT by employing the full-potential linearized augmented plane wave code
WIEN97 \cite{WIEN97} and by LMTO \cite{Andersen_75} based on the 
Stuttgart TBLMTO-47 code. We find complete agreement in between these two
calculations.

The overall band picture for $\gamma$-LiV$_2$O$_5$ (Fig.\ \ref{Li_spag})
is similar to that of $\alpha '$-NaV$_2$O$_5$ \cite{Smolinski98}. 
The V-$3d$ states give the predominant contribution to the bands at the
Fermi level and up to $\approx 4\mbox{eV}$ above it. The lower valence
bands are mainly O-$2p$ states and are separated by a gap of 
$\approx 2.2\,\mbox{eV}$ from the bottom of the V-$3d$ bands. 
There is  a nonequivalent contribution
from the two types of V-sites, V(1) and V(2). 
The four lowest-lying $3d$ bands at the  Fermi level 
 are half-filled \cite{clarification} and are made
up predominantly of V(1)-$3d$ and of V(2)-$3d$ 
states in the ratio $p(1)/p(2)\approx 2:1$ and $3:1$ depending
on the $k$-values.
The next four bands less than $1\,\mbox{eV}$ above
the Fermi level also exhibit a mixture of 
V(1)-$3d$ and V(2)-$3d$ character.

\begin{figure}[t]
\centerline{
\epsfig{file=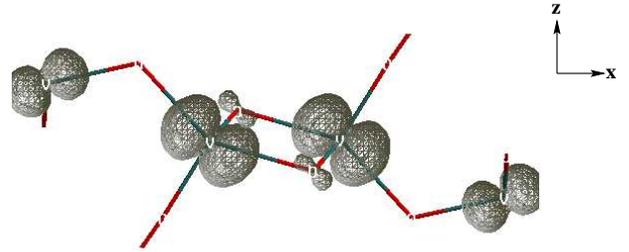,width=0.45\textwidth} %,angle=-90}
           }
\vspace{4pt}
\caption{\label{Li_ED}
Electron density of the V-$3d$ bands seen in a $(xz)$ cut. Shown is
 the isoline 0.05 e/$\AA^3$. The bigger lobes correspond to V(1)-$3d$ orbitals
and the smaller ones to V(2)-$3d$ orbitals. Note that the tilted
orbitals point 
 towards the bridging oxygens (compare with Fig.\ \protect\ref{struc_xz}).
}
\end{figure}

The vanadium bands at the Fermi-level are
of d$_{xy}$  symmetry (global symmetry)
with a certain admixture with the
d$_{yz}$ state due to the rotation of the 
basal-plane of the V(1/2)O$_5$ pyramids 
with respect to the $x$-axis (see 
Fig.\ \ref{struc_xz}).  The
 degree of admixture for both vanadium types is such that
the respective V-$3d$ orbitals point -as in
$\alpha'$-NaV$_2$O$_5$ \cite{Smolinski98}- 
roughly towards the bridging oxygens. In this sense,
one can regard the electronic active
V(1)-$3d$ and V(2)-$3d$ orbitals as $d_{xy}$ orbitals rotated
around the $y$-axis by angles 
 $\varphi_{1}=35^\circ$ and
$\varphi_{2}=-28^\circ$ respectively (see  Fig.\ \ref{Li_ED} ).

The most notable difference  between the band-structure of
$\gamma$-LiV$_2$O$_5$ and 
$\alpha '$-NaV$_2$O$_5$ is the band-splitting at the
X and T points  in $\gamma$-LiV$_2$O$_5$, which is absent
in $\alpha '$-NaV$_2$O$_5$. This splitting is due to the existence of
two different V-sites in $\gamma$-LiV$_2$O$_5$ as we will
see in the next paragraph. The fact that this 
splitting is big, close to the overall bandwidth,
indicates already that the microscopic parameters associated
with the V(1) and V(2) sites must differ substantially.

 Also note
 that the splitting of the bands
at the Fermi-level due to the existence of two $xy$-planes 
in the crystallographic unit-cell of $\gamma$-LiV$_2$O$_5$
is small and does not occur along the path Z-U-R-T-Z. 
We will concentrate
upon the discussion of the in-plane dispersion in what follows.

%%%%%%%%%%%%%%%%%%%%%%%%%%%%%%%%%%%%%%%%%%%%%%%%%%%%%%%%%%%%%%%%%%%
%%%%%%%%%%%%%%%%%%%%%%%%%%%%%%%%%%%%%%%%%%%%%%%%%%%%%%%%%%%%%%%%%%%

{\it Microscopic parameters.-}
In order to determine the microscopic model 
appropriate for $\gamma$-LiV$_2$O$_5$ we have analyzed
the band-structure shown in Fig.\ \ref{Li_spag} by
a (minimal) tight-binding model with
one orbital per vanadium site,
which generalizes the tight-binding model appropriate
for $\alpha'$-NaV$_2$O$_5$ to the case of two different
V-sites. A straightforward fit, e.g.\ by least-squares, 
is not possible  since the lower unoccupied 
V-bands (roughly in between $0.5-1.0\,\mbox{eV}$), are strongly
hybridized with the O-$p$ orbitals and in order to describe
the low-energy physics of this system these bands should also
 be considered.  

 In recent years \cite{Andersen00}, a new version of the
LMTO method has been proposed and implemented which is proved to be
powerful in providing an effective orbital representation 
of the system by integrating out the higher degrees of 
freedom using the so-called
downfolding technique.
The usefulness of the method lies in taking
into account proper renormalization effects. However, the Fourier
transform of the downfolded Hamiltonian to extract the tight-binding
parameters, results  in long-ranged
hopping matrix elements. 

\begin{figure}[t]
\centerline{
\epsfig{file=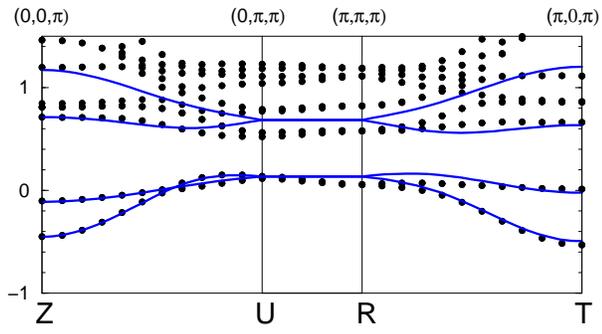,width=0.45\textwidth} %,angle=-90}
           }
\vspace{4pt}
\caption{\label{band_fit}
Comparison of the tight-binding fit (solid lines)
with the DFT bands (filled circles). The parameters
used (see Fig.\ \protect\ref{band_Li}) are (in eV)
$\epsilon_0=0.15$, $t_a=0.35$, $t_d=0.10$,
$t_b^{(1)}=-0.06$, $t_b^{(2)}= 0.02$,
$t_1^{(1)}=-0.18$, $t_2^{(1)}=-0.05$,
$t_1^{(2)}=0.05$, $t_2^{(2)}=-0.02$.
}
\end{figure}

We have therefore considered a combination of both methods. We
use only the short-ranged hopping matrix elements 
(see Fig.\ \ref{band_Li}) provided by
the downfolding procedure as an input
 for the tight-binding model. These
matrix elements are then optimized to reproduce the behavior of the
 {\it ab-initio} bands near the Fermi level.
The result and the parameters (apart from an overall
constant energy) of the optimal fit
are shown in Fig.\ \ref{band_fit}.

From the relative weight between the V(1) and the 
V(2) contributions near the Fermi level,
$p(1)/p(2)$, we learn that 
there must be a substantial on-site energy 
$\pm\epsilon_0$ for the
$V(1/2)$ orbitals respectively. 
We find $\epsilon_0=0.15\,\mbox{eV}$.
The rung-hopping matrix element $t_a$ may be expected,
on the other hand, to be quite close to the
one obtained for $\alpha '$-NaV$_2$O$_5$\cite{parameters_Na}
since the large bending angle V(1)-O(1)-V(2) should
not substantially affect the $\pi$-bonding via the O(1)-$p_y$ orbital.
Indeed, we find $t_a=0.35\,\mbox{eV}$. We can check whether
our estimates for $\epsilon_0$ and $t_a$ lead to the correct
V(1)-V(2) charge ordering. By diagonalizing a simple
two-site rung model we 
find the relation
\begin{equation}
{e_0\over t_a}\ =\ {p(2)-p(1)\over 2\sqrt{p(1)p(2)}}~,
\label{rung}
\end{equation}
which yields $p(1)/p(2)=2.3$ for $e_0/t_a=0.15/0.35$,
i.e\ $p(1)\approx0.7$ and $p(2)\approx0.3$, in
agreement with the DFT results.

We note next, that the crossing along $\Gamma$-Y and Z-U respectively
can be described (within a V-model)
only by a considerable $t_2^{(1/2)}$, as it has been
 noted previously \cite{Smolinski98}. Recently
Yaresko {\it et al.} \cite{Yaresko00} have discussed
that this matrix element arises naturally when one 
integrates out the coupling  between two leg-oxygens
of two adjacent edge-sharing VO$_5$ pyramids.
Our results $t_2^{(1)}=-0.05\,\mbox{eV}$ and
$t_2^{(2)}=-0.02\,\mbox{eV}$ are close to the values
obtained for $\alpha '$-NaV$_2$O$_5$\cite{parameters_Na}.

A substantial diagonal hopping matrix element $t_d$ 
is needed in order to explain the fact that
the dispersion along $\Gamma$-Y of the lower four V-bands 
has an opposite behavior with respect 
to the upper four V-bands \cite{explain_td}, 
as has been pointed out independently for 
$\alpha '$-NaV$_2$O$_5$ by 
Yaresko {\it et al.} \cite{Yaresko00}.
The substantial contribution of $t_d=0.10\,\mbox{eV}$ 
can be explained by effective V-O-O-V exchange paths 
\cite{Yaresko00,Smolinski_diss}. 
The hopping matrix element along the V(1) leg, 
$t_b^{(1)}=-0.06\,\mbox{eV}$ has a similar value to the one for
$\alpha'$-NaV$_2$O$_5$, though
the sign of the effective V(2)-leg hopping
parameter, $t_b^{(2)}=0.02\,\mbox{eV}$, is opposite to
the expected one. 

The result for $t_1^{(2)}=0.05\,\mbox{eV}$ might have
been expected since its corresponding value for
$\alpha '$-NaV$_2$O$_5$ is small\cite{parameters_Na}.
The result $t_1^{(1)}=-0.18\,\mbox{eV}$ is, on the other
hand, substantially larger and needs some explanation. 
In terms of the bandstructure, $t_1^{(1)}$ is  determined
predominantly by the large splitting of the V-d bands \cite{dispersion_X}
at X and T (see Figs.\ \ref{Li_spag} and \ref{band_fit}).  Also,
it has been noted previously \cite{Smolinski98,Horsch98} in the context
of $\alpha '$-NaV$_2$O$_5$
that the bare two-center Slater-Koster matrix elements contributing
to $t_1$ can be as large as $-0.3\,\mbox{eV}$.
The effective $t_1$ is reduced from the bare 
two-center matrix element by the rotation of the
V-$3d$ orbitals about the crystallographic $y$-axis and by
interference from 
three-center terms\cite{Smolinski98,Smolinski_diss}.
The contribution from V-O-V exchange paths depends 
strongly on the relative positions of the three atoms.
We have performed for $\gamma$-LiV$_2$O$_5$ a Slater-Koster analysis 
and found that due to interference effects in between
the respective $dd\sigma$ and $dd\pi$ contributions,
 the  V(2)-V(2) matrix element contributing 
to $t_1^{(2)}$ is smaller than
 the V(1)-V(1) matrix element contributing to
$t_1^{(1)}$.  In addition,
the exchange along V(1)-O(4)-V(1) contributing
to $t_1^{(1)}$ was found to be substantially larger
than the exchange
V(2)-O(5)-V(2) contributing to $t_1^{(2)}$.  
The particular V-O-V distances and angles lead therefore
to different $t_1^{(1)}$ and $t_1^{(2)}$.

%%%%%%%%%%%%%%%%%%%%%%%%%%%%%%%%%%%%%%%%%%%%%%%%%%%%%%%%%%%%%%%%%%%
%%%%%%%%%%%%%%%%%%%%%%%%%%%%%%%%%%%%%%%%%%%%%%%%%%%%%%%%%%%%%%%%%%%

{\it Microscopic model.-}
The microscopic model corresponding to the
results shown in Fig.\ \ref{band_fit}
is that of a spin-$\frac{1}{2}$ antiferromagnetic 
Heisenberg chain where the magnetic moments
are associated dominantly with the sites of the
V(1) ions in $\gamma$-LiV$_2$O$_5$, the contribution
is $p(1)\approx0.7$.  The contribution
of about $p(2)\approx0.3$ of the V(2) to the magnetic moment on a 
$V(2)-O(1)-V(1)$ rung has nevertheless important
consequences for the underlying microscopic model.
For negligible values of $p(2)$ the microscopic model
could be considered as that of a zig-zag chain with 
 a $J \sim 4 \frac{\left(t_1^{(1)}\right)^2}{U}$
 since
$t_b^{(1)}=-0.06\,\mbox{eV}$ is smaller than 
$t_1^{(1)}\approx-0.18\,\mbox{eV}$.  In the presence of a non-negligible
value of $p(2)$ 
the effective hopping matrix element $t_b^{(eff)}$
in between two asymmetric rung states along $b$ is
\[
t_b^{(eff)} = p(1)\,t_b^{(1)}
              + p(2)\,t_b^{(2)} - 2
\sqrt{p(1)p(2)}\,t_d\ \approx\ -0.127\,\mbox{eV}
\]
suggesting an asymmetric ladder model.
Then, using the expression $J_b=2 \frac{\left(t_b^{(eff)}\right)^2}{E_c}$,
and assuming that the charge-transfer gap is $E_c\approx0.7\,\mbox{eV}$ as
in $\alpha'$-NaV$_2$O$_5$\cite{Smirnov_98}
the exchange integral is $J_b \approx 540\,\mbox{K}$ which overestimates 
the experimental value $J_{exp}=308\,\mbox{K}$. The
degree of charge ordering has therefore a substantial
influence on the nature of the magnetic couplings \cite{note_t1}.

%%%%%%%%%%%%%%%%%%%%%%%%%%%%%%%%%%%%%%%%%%%%%%%%%%%%%%%%%%%%%%%%%%%
%%%%%%%%%%%%%%%%%%%%%%%%%%%%%%%%%%%%%%%%%%%%%%%%%%%%%%%%%%%%%%%%%%%

{\it Conclusions.-}
We have presented an analysis of DFT band-structure
calculations for $\gamma$-LiV$_2$O$_5$. We find that the
degree of charge ordering has a substantial influence
on the nature of the magnetic state. Our results indicate
incomplete charge ordering and  $\gamma$-LiV$_2$O$_5$ could in this case
be viewed 
 as a spin-$\frac{1}{2}$ asymmetric quarter-filled ladder
compound.  This model
 would explain the spin wave excitation
spectrum obtained by inelastic neutron scattering experiments \cite{Takeo_99}.
Finally, we observe that small distortions in the lattice 
may have substantial effects on the interladder
V-V hopping matrix element $t_1$.

{\it Acknowledgments.-}
 This work was partially supported by the DFG.
  One of us (R.V.) would like to thank C.O. Rodriguez for helpful advice 
 regarding the WIEN97 code and A. Kokalj for providing the graphics
XCrysDen code.

%%%%%%%%%%%%%%%%%%%%%%%%%%%%%%%%%%%%%%%%%%%%%%%%%%%%%%%%%%%%%5
%%%%%%%%%%%%%%%%%%%%%%%%%%%%%%%%%%%%%%%%%%%%%%%%%%%%%%%%%%%%%5

%%%%%%%%%%%%%%%%%%%%%%%%%%%%%%%%%%%%%%%%%%%%%%%%%%%%%%%%%%%%%5
%%%%%%%%%%%%%%%%%%%%%%%%%%%%%%%%%%%%%%%%%%%%%%%%%%%%%%%%%%%%%5
%\newpage

%    ------------------                                                  
%    ------------------                                                  

%    ------------------                                                  
%    ------------------                                                  

%    ------------------                                                  
%    ------------------                                                  

%    ------------------                                            
%    ------------------                                                  

%    ------------------                                                  
%    ------------------                                                  
   
%    ------------------                     

\begin{thebibliography}{99}
\bibitem{clarification} Note that the unit cell of LiV$_2$O$_5$ corresponds
                        to four formula units.
\bibitem{explain_td} The dispersion of the antibonding/bonding
    V-bands in $\alpha '$-NaV$_2$O$_5$ is 
    $\pm t_a + 2(t_b\pm t_d)\cos(k_y)$. In order to obtain
    an opposite 
    dispersion\protect\cite{Smolinski98,Yaresko00} one
    needs $t_d > |t_b|$. The bandwidth 
    $0.7\,\mbox{eV}\approx4|t_b-t_d|$ of the bonding orbital 
    and the smaller bandwith $\approx0.1\,\mbox{eV}$ 
    of the antibonding orbital yield then
    $t_d\approx 0.095\,\mbox{eV}$ and
    $t_b\approx-0.08\,\mbox{eV}$.
\bibitem{parameters_Na} The tight-binding parameters for
   $\alpha '$-NaV$_2$O$_5$ 
   are\protect\cite{Smolinski98,Yaresko00,explain_td}
   $t_a=0.375\,\mbox{eV}$,
   $t_d=0.095\,\mbox{eV}$, $t_b=-0.08\,\mbox{eV}$,
   $t_2=-0.055\,\mbox{eV}$.
   We have reanalyzed the DFT band-structure of
   NaV$_2$O$_5$ and found that the effective $t_1$  is
   very small.
\bibitem{dispersion_X} For general parameters the tight-binding
                       equation cannot be solved analytically but
 certain limiting cases can be studied. As a illustration, 
       we note that for $t_1^{(1)}=
                         t_1^{(2)}=t_1$ $t_b^{(2)}=-t_b^{(1)}$
    and $\epsilon_0^{(2)}= \epsilon_0^{(1)}$
                      the dispersion of the antibonding/bonding bands
                          at X and T in $\gamma$-LiV$_2$O$_5$
                        is given by $\lambda ^2 =(\epsilon_0 +
    (t_b^{(1)}-t_b^{(2)})\pm 2t_1)^2 +(t_a+2t_d)^2$.
\end{thebibliography}
\end{document}